# Astronomy at the University of Salamanca at the end of the 15th century

## What *El Cielo de Salamanca* tells us

Carlos Tejero Prieto

Departamento de Matemáticas, Instituto de Física Fundamental y Matemáticas

Universidad de Salamanca, Plaza de la Merced 1-4, 37008, Salamanca, Spain

e-mail: carlost@usal.es

Abstract: *El Cielo de Salamanca* ("The Sky of Salamanca") is a quarter-sphere-shaped vault 8.70 metres in diameter. It was painted sometime between 1480 and 1493 and shows five zodiacal constellations, three boreal and six austral. The Sun and Mercury are also represented. It formed part of a three times larger vault depicting the 48 Ptolemaic constellations and the rest of the planets known at the time. This was a splendid work of art that covered the ceiling of the first library of the University of Salamanca, one of the oldest in Europe having obtained its royal charter in 1218. But it was also a pioneering scientific work: a planetarium used to teach astronomy, the first of its kind in the history of Astronomy that has come down to us in the preserved part that we now call "The Sky of Salamanca". We describe the scientific context surrounding the chair of Astrology founded around 1460 at the University of Salamanca, which led to the production of this unique scientific work of art and to the flourishing of Astronomy in Salamanca. We analyse the possible dates compatible with it, showing that they are extremely infrequent. In the period of 1100 years from 1200 to 2300 that we studied there are only 23 years that have feasible days. We conclude that the information contained in *El Cielo de Salamanca* is not sufficient to assign it to a specific date but rather to an interval of several days that circumstantial evidence seems to place in August 1475. The same configuration of the sky will be observable, for the first time in 141 years, from the 22nd to the 25th of August 2022. The next occasion to observe it live will be in 2060.

**A unique work of art**

At the back of the Plateresque Patio de Escuelas Menores of the University of Salamanca, there is a building that houses a work of art unique in its beauty, its subject matter and its importance for the history of Astronomy. On the door, painted in red on Villamayor sandstone, the visitor is told that it is *El Cielo de Salamanca* ("The Sky of Salamanca"). On the lintel there is a commemorative plaque dedicated to the 17th century playwright, poet and writer Calderón de la Barca on his second centenary. After passing through the entrance, we find a modern sliding smoked glass door that opens automatically as we approach and protects the interior of the building from excessive light coming in from the outside. It leads to a first section of corridor in semi-darkness. Walking along it gives visitors' eyes time to adjust to the splendid work of art that they will find soon afterwards on their right in a dimly lit room, in accordance with what they are about to contemplate: the representation of a starry night sky.

Little by little, our eyes finish adapting and before us appear the delicate blue tones of the background, the gold-painted stars and the mythological images of various constellations together with two planets (etym. Greek πλανήτες; properly 'wandering', because they move against the background of fixed stars) the Sun and Mercury, who are depicted triumphant on chariots, drawn by four horses in the case of the Sun and by two eagles in the case of swift Mercury. The Sun and the Moon were at the time considered as planets.

We can see painted five zodiacal constellations (Leo, Virgo, Libra, Scorpio and Sagittarius), three northern constellations (Boötes, Hercules and Ophiuchus) and six southern constellations (Hydra, Crater, Corvus, Ara, Centaurus and Corona Australis). In the lower part of the vault, four winds are represented: Zephyr, Austro, Eurus and Boreas. As a picture is worth a thousand words, we can contemplate it in Figs. 1 and 2.

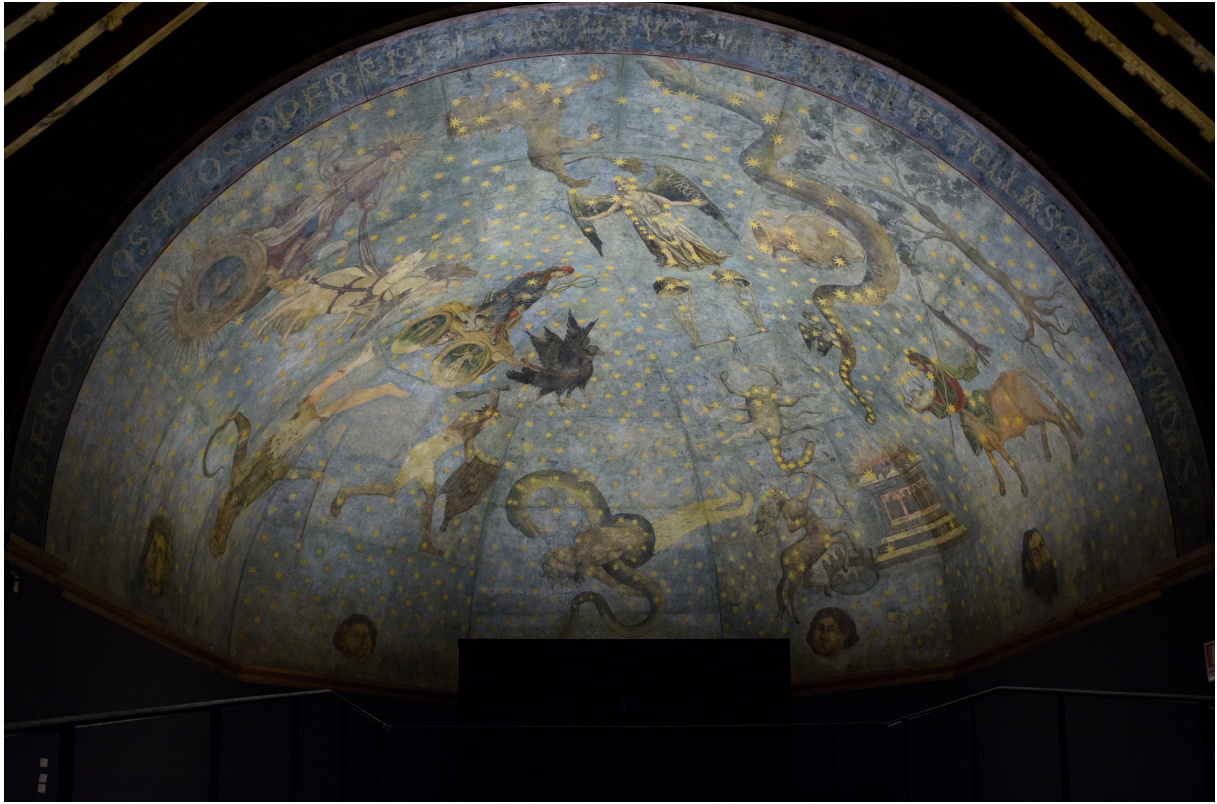

*Fig. 1. "The Sky of Salamanca" as it can be seen today in the Patio de Escuelas Menores. USAL*

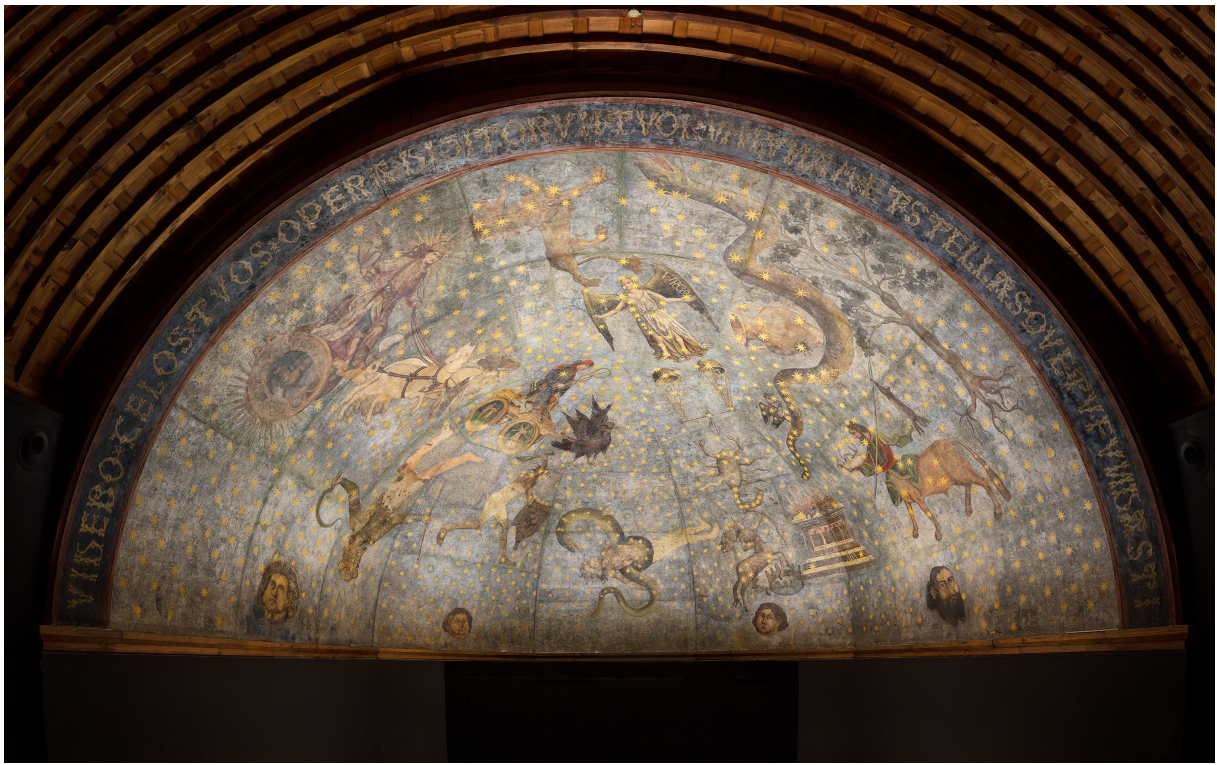

*Fig 2. . "The Sky of Salamanca" under a higher illumination. USAL*

*El Cielo de Salamanca* was not always in its current location. It was originally part of the first library of the University of Salamanca, built between 1474-1479 in the site now occupied by the Chapel of San Jerónimo in the Edificio de Escuelas Mayores [5, pp. 30 and 33]. The execution of the pictorial work on the vault of the library took place sometime between 1480 and 1493 [5, p. 40]. According to specialists, it was painted by the famous Salamancan painter Fernando Gallego (1440-1507), who belonged to the Hispano-Flemish Gothic School.

However, the "Sky of Salamanca", both for its subject matter, analogous to that of certain Italian works of the same period, and for its grandiose proportions — it is a quarter of a sphere 8.70 metres in diameter —, can be considered, as Martínez Frías has pointed out, [7, p. 115], a precedent of the Renaissance style.

What we can see today is approximately one third of the vault decorated by Gallego. The rest collapsed during a renovation carried out in the 18th century. The undamaged part of the vault was hidden in the subsequent restoration works. It was rediscovered in 1901 by Professor García Boiza [5, p. 55]. In 1950 the paintings were removed from the original vault using the strappo technique. After an arduous restoration process, they were transferred in 1953 to their present location. For more details see Chapter IV of the comprehensive and scholarly monograph [5].

After an obligatory break due to the restoration of the Escuelas Menores, on 3 December 2021 the visits to *El Cielo de Salamanca* resumed. It is therefore a good time to enjoy this work of art live again. If this is not possible, there is always the option of a virtual visit, which can be accessed in the following [link](link).

From the description made by Diego Pérez de Mesa in 1590 [5, p. 26], we know that the original vault of the ancient library contained the 48 Ptolemaic constellations. These were formed by the 12 zodiacal constellations that define the ecliptic, together with the 21 boreal and 15 austral constellations, located to the north and south of the ecliptic, respectively. The presence of the Sun and Mercury in *El Cielo de Salamanca*, together with the descriptions of the library made by Hyeronimus Münzer and Lucio Marineo Sículo in the 90s of the 15th century, allow us to infer that the original vault also contained the rest of the planets known at that time: the Moon, Venus, Mars, Jupiter and Saturn, [6, p. 114].

Since its rediscovery, several authors have wondered whether *El Cielo de Salamanca* corresponds to a specific date and, if so, whether it was chosen for a specific reason. Before giving our answer, it is convenient to put in context the origin of this vault.

**Why was this type of representation chosen?**

In the second half of the 15th century, Astronomy flourished at the University of Salamanca. We know that around 1460 a chair of Astrology was created. It was occupied by Nicolás Polonio, probably of Polish origin, until 1464. It should be noted that at that time Astrology was one of the seven liberal arts and was synonymous with Astronomy [4, p. 118]. Astrology was then considered indispensable for the practice of medicine, as it was believed that in order to cure their patients, doctors had to know their horoscopes [4].

We know that Polonio, taking 1460 as his root, adapted to the Salamanca meridian the Tabulae Resolutae. These are a version of the Alphonsine tables drawn up by Andreas Grzymala around 1449 for the students of the University of Kraków. They differ significantly from the Paris version of the Alphonsine tables. The mean movements are presented in cycles of 20 years and the zodiacal signs are 30° wide arcs unlike in the Parisian version.

After the arrival of Nicolás Polonio in Salamanca, the chair of Astrology began to develop an intense activity, playing an important role in the remainder of the 15th century and throughout the 16th century. Witness to this is the abundant written output of its professors and former students. They produced astronomical material specifically for Salamanca. It is also corroborated by the fact that the General Library of the University has many collections related to Astronomy, second only to those dedicated to Grammar and Poetic Rhetoric.

The list of holders of the chair of Astrology in the 15th century is as follows:

> (1460 -1464) Nicolás Polonio
>
> (1464-1469) Juan de Selaya
>
> (1469-1475) Diego Ortiz de Calzadilla
>
> (1476-1480) Fernando de Fontiveros
>
> (1481-1495) Diego de Torres
>
> (1495-1504) Rodrigo de Vasurto

The chair of Astrology during the 15th century was practically monopolised by the fellow students of Colegio Mayor de San Bartolomé, which had been created in 1401. With the exception of Nicolás Polonio and Diego Torres, the rest were Bartolomeans.

The development of Astronomy at that time did not only take place at the university. In 1452 Abraham Zacut was born in Salamanca and would eventually become the most important astronomer in the Iberian Peninsula during the last quarter of the 15th century. Between 1478 and 1479 he wrote the "Ha Hibbur Ha Gadol", "The Magna Compilation". Its Root is 1473. It contains 65 Tables adapted to the Salamanca meridian and 19 chapters of canons. Zacut's tables are largely dependent on the Tabulae Resolutae of Nicolás Polonio, using the same conventions of 20-year cycles and natural zodiacal signs of 30°.

We know that Zacut was related to Juan de Selaya, the professor of astrology who succeeded Polonio. In fact, incunabulum I 176 of the library of the University of Salamanca contains a manuscript with the Spanish translation of the "Ha Hibbur Ha Gadol". In the Explicit we are told:

> *"This book was translated from Hebrew into Spanish in the year 1481 by Master de Selaya through Abraham Zacut, as interpreter."*

However, Zacut is best known for his Almanach Perpetuum published in Leiría, Portugal, in 1496. It is based in part on the "Ha Hibbur Ha Gadol", although they are distinct works. It consists of 23 chapters with the canons and 52 astronomical tables. It contains the tables of solar positions. These would become the basis for the Portuguese nautical tables of the 16th century.

Unquestionable proof of the importance that the chair of Astrology had acquired can be seen in the fact that, once the construction of the University library was completed, it was decided that its vault should be decorated with a representation of the constellations, planets and stars.

**A pioneering scientific work**

However, the decoration of the vault of the first library of the University of Salamanca would not only be a work of art. As Flórez Miguel has pointed out, [3, p. 187,] it was a scientific work, a planetarium painted on the ceiling of the library whose first and foremost aim was the practical teaching of astrology. This planetarium is thus the first of its kind in the history of Astronomy to have come down to us in the part we now call *El Cielo de Salamanca*. It should be noted that, given that the surface of the original vault was a half cylinder topped by two quarters of a sphere, the representation of celestial objects on it was not particularly simple. Therefore, we can think that what was painted on it was a qualitative view of the night sky.

At that time the position of the zodiacal constellations was well known. Recall that these are the constellations which, when seen from the Earth, are approximately in the same plane as the ecliptic. The ecliptic was defined as the curved line along which the Sun moved around the Earth in its "apparent annual motion". Solar eclipses occur along it, hence the name. Stars were assumed to be fixed on a geocentric sphere, and the Sun was said to move through the year along the zodiacal constellations. Copernicus (1473-1543) would be the one to put the Sun at the centre of the Universe, but that was long after the "Sky of Salamanca" was painted. Even so, the plane of the ecliptic remains the same in the geocentric model as in the heliocentric model. According to this the zodiacal constellations should make a complete cycle annually. This can be taken as correct if we refer to a time span of a few centuries. However, this regularity is lost when we consider longer time spans, due to the so-called precession of the equinoxes. This is caused by the change of orientation in space of the Earth's rotation axis with a period of about 25800 years. Surprisingly, this was already observed by Hipparchus of Nicaea in the 2nd century B.C. This same movement is the reason why today zodiacal signs do not correspond to zodiacal constellations.

Fernando Gallego probably followed the instructions of expert astronomers, since the constellations are represented in their real order. However, as we have already said, it is not a scale representation like the one contained in today's planispheres. In any case, the relative positions of the constellations do approximate the real situation of the constellations. However, only some of the painted stars correspond to visible stars. Others seem to have been introduced by the author for aesthetic reasons. In the "Sky of Salamanca" the only two planets that appear are the Sun and Mercury. The former is painted in the constellation of Leo, while the latter is in Virgo, just below the Sun.

**A Midsummer Night's Sky**

Since it was rediscovered, several authors have wondered whether the position of the heavenly bodies in *El Cielo de Salamanca* corresponds to a specific day. It is not easy to give a conclusive answer. In fact, based exclusively on what we see painted there, it is literally impossible to give a definitive answer as, due to the periodicity of the orbits of the planets, there are innumerable dates on which a situation like the one depicted could occur.

What is certain is that what is depicted on the vault is the position of the constellations in the month of August, which is when the Sun is in the constellation Leo. The presence of Mercury and the absence of the other planets known at the time (Venus, Mars, Jupiter, Saturn and the Moon) impose additional conditions, but even so we still have countless possibilities.

It is therefore necessary to introduce some additional hypothesis to limit the feasible cases. The approach of the authors who have analysed this question has been to limit the search for possibilities to a certain period of time.

To analyse the solutions previously provided in the literature, we have made use of the free and open source astronomy software Stellarium v0.21.3 which can be downloaded from the following [link](). This software performs graphical representations implementing the VSOP87 method for the calculation of planetary ephemerides. This mathematical model was developed by P. Bretagnon and G. Francou [1] at the Bureau des Longitudes in Paris in order to describe the long-term variations of planetary orbits. It guarantees for Mercury, Venus, the Earth-Moon barycentre and Mars an accuracy of 1" for 4000 years before and after the J2000 epoch. The same accuracy is given for Jupiter and Saturn for 2000 years before and after J2000. In addition, as a double check, we have verified the results obtained using the Horizons system of NASA's Jet Propulsion Laboratory. This represents the state of the art in ephemeris calculation and is accessible at the link. All calculations are expressed in UTC time (Coordinated Universal Time). The constellations are considered according to the IAU (International Astronomical Union) definition.

We will now proceed to analyse in chronological order all the proposals that we have been able to find in the literature.

The first author we know of to address this question is the astronomer Ernst Zinner, director of the Remeis Observatory in Bamberg, Germany, between 1926 and 1956. His main speciality was Renaissance Astronomy and the History of Astronomical Instruments. In 1960, in the article [10, p. 10], he considered the period 1473-1493, concluding that the "Sky of Salamanca" represented the night of 6 August 1475. The author does not tell us what means he used to reach this conclusion. The calculation we have made with Stellarium v0.21.3 for that night produces the results shown in Fig. 3, from which it follows that it is not a feasible solution for two reasons. Firstly, because Mercury is in Leo and secondly because the Moon is in Virgo, so it should be represented in the "Salamanca Sky".

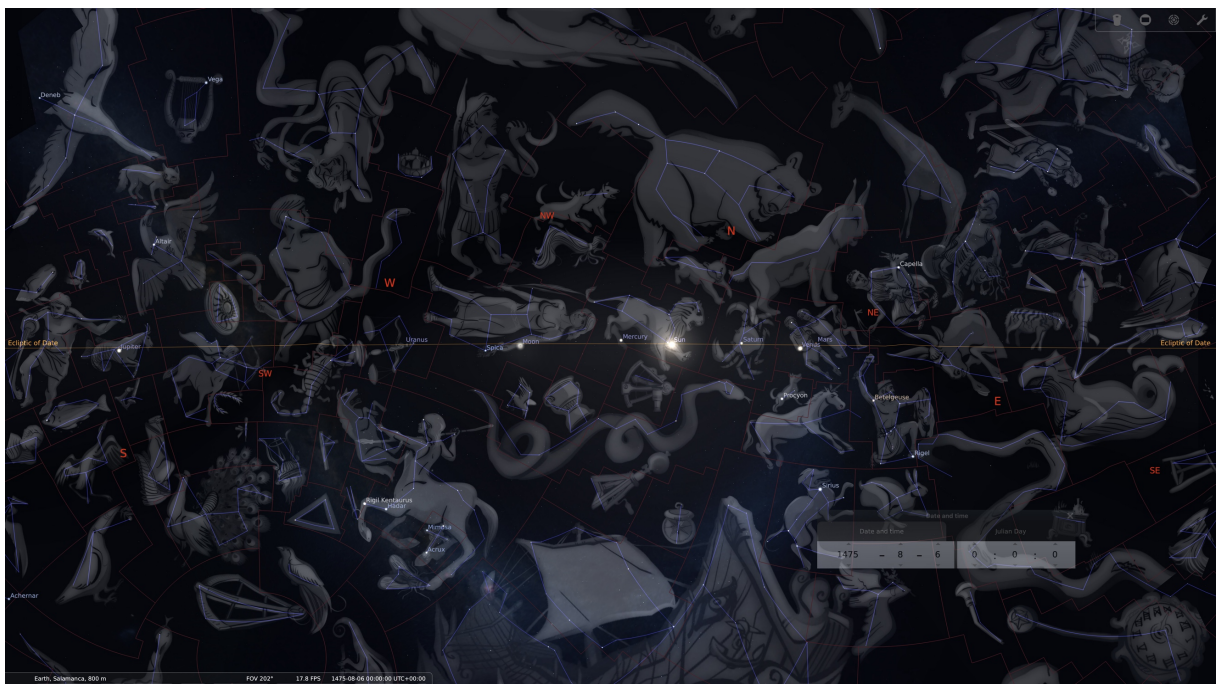

*Fig. 3. Celestial sphere plotted with Stellarium v0.21.3 in Mercator projection at 0:0:0 UTC on 6 August 1475.*

The next work was the exhaustive study by Professor Noehles-Doerk in 1992 on the University Library of Salamanca in the 15th century and its cosmological painting programme. In it [8 pp. 26-27], she states [our translations from German]:

> "If one starts from a memorable event from the time of the planning or construction of the library in Salamanca, then according to the position of the planets in the signs of the zodiac, with the absence of the moon in the depicted configuration, it can only be a date in August between the 14th and 29th of the year 1475."

She herself explains how she came to this conclusion [8, p. 27]:

> "I owe this result to the Münster astronomer Hilmar W. Duerbeck."

She also gives us a summary of the astronomer's reasoning:

*"In the period 1474 - 1494, the represented configuration of Sun in Leo, Mercury in Virgo, while the planets Venus, Mars, Jupiter and Saturn are not in the zodiacs Leo, Virgo, Libra and Sagittarius, is fulfilled only in August 1475. The Sun is in Leo every August, but Mercury can be, for example, in Cancer, Leo or Virgo: hence years in which Mercury is in Leo can be crossed off the list of eligible years. Similarly, Venus may be in the constellations of Gemini, Cancer, Leo, Virgo or Libra in August; obviously, it was in the unpreserved constellations of Gemini or Cancer at the time of the sky depicted. According to this method, which uses the non-visibility of the planets Venus, Mars, Jupiter and Saturn to exclude individual years, the only one remaining is the year 1475: the position of the planets (on 20.8.1475) is as follows:*

|  | *Ecliptic longitude* | *Constellation* | *Observation* |
|---|---|---|---|
| *Sun* | 158° | *Leo* | *present* |
| *Mercury* | 180° | *Virgo* | *present* |
| *Venus* | 121° | *Cancer* | *not present* |
| *Mars* | 109° | *Gemini* | *not present* |
| *Jupiter* | 302° | *Capricorn* | *not present* |
| *Saturn* | 112° | *Cancer* | *not present* |

*Since the Moon, as the "seventh planet", was also depicted and the time - due to its rapid movement through the zodiac - can be accurately determined to about 2 days, we can name periods when it did not appear in the depicted images: July to September 1475. The best time interval to consider is 14 to 29 August 1475, when the Moon passes through ecliptic longitudes 289° to 135° (the constellations Capricorn, Aquarius, Pisces, Aries, Taurus, Gemini or Cancer)."*

Since the summary of Duerbeck's work by Noehles-Doerk does not mention which epoch and reference system is used (ecliptic of date or ecliptic B1950, or other), we have repeated the calculations for the period under consideration. Our results essentially agree with those of Duerbeck, except that 14 and 29 August 1475 must be excluded as shown in Figs. 4 and 5. The first of these dates is not correct because the Moon was in Sagittarius, so we should see it in the "Sky of Salamanca". The second case is not possible since the Moon was in Leo.

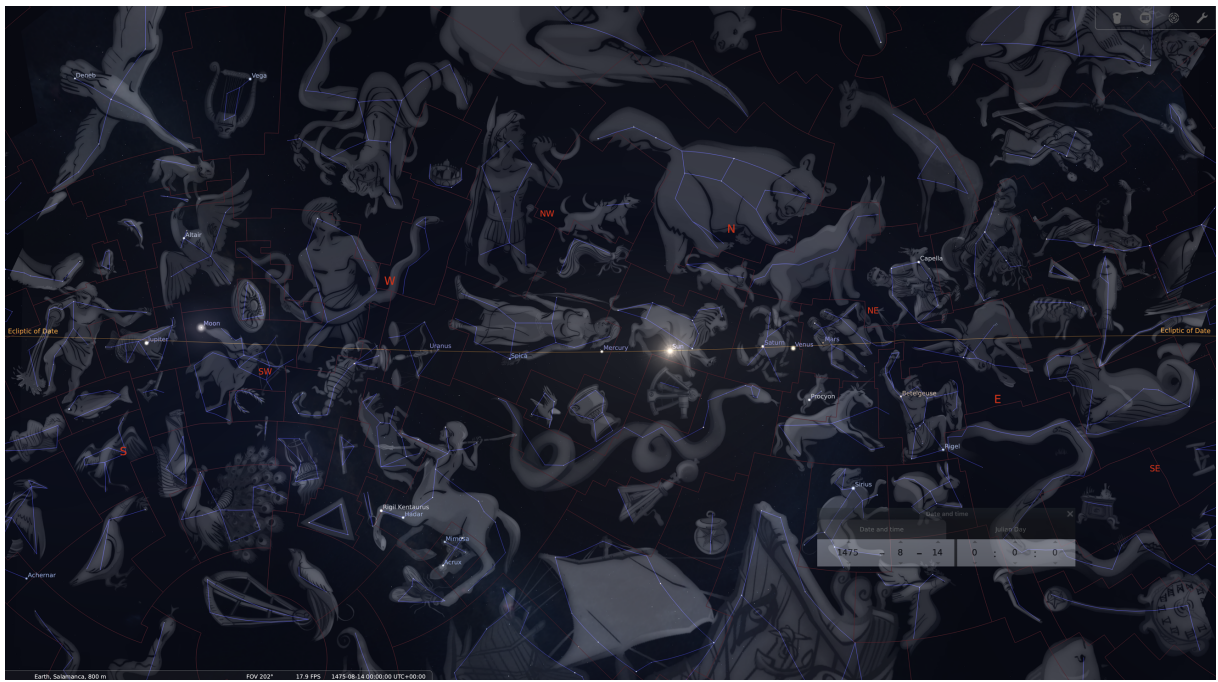

*Fig. 4. Celestial sphere plotted with Stellarium v0.21.3 in Mercator projection at 0:0:0 UTC, 14 August 1475.*

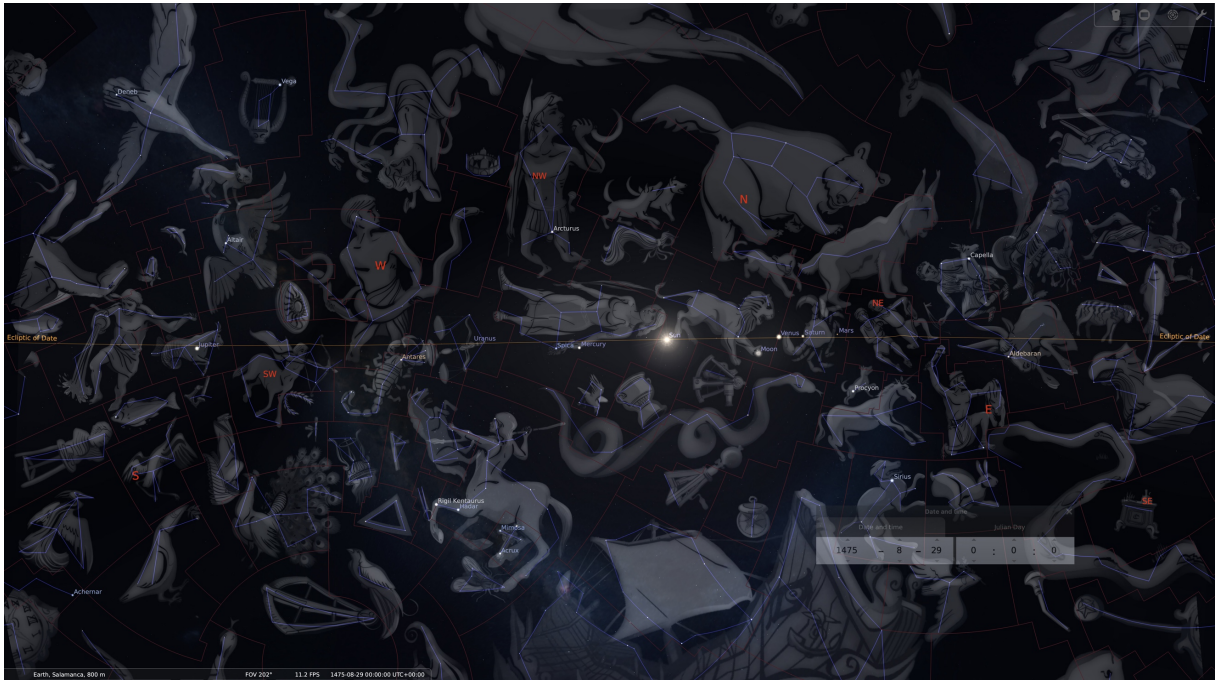

*Fig. 5. Celestial sphere plotted with Stellarium v0.21.3 in Mercator projection at 0:0:0 UTC, 29 August 1475.*

On the other hand, the period between the 15th and the 28th August 1475 is feasible. Figs. 6 and 7 show the calculations made for the initial and final days of this interval.

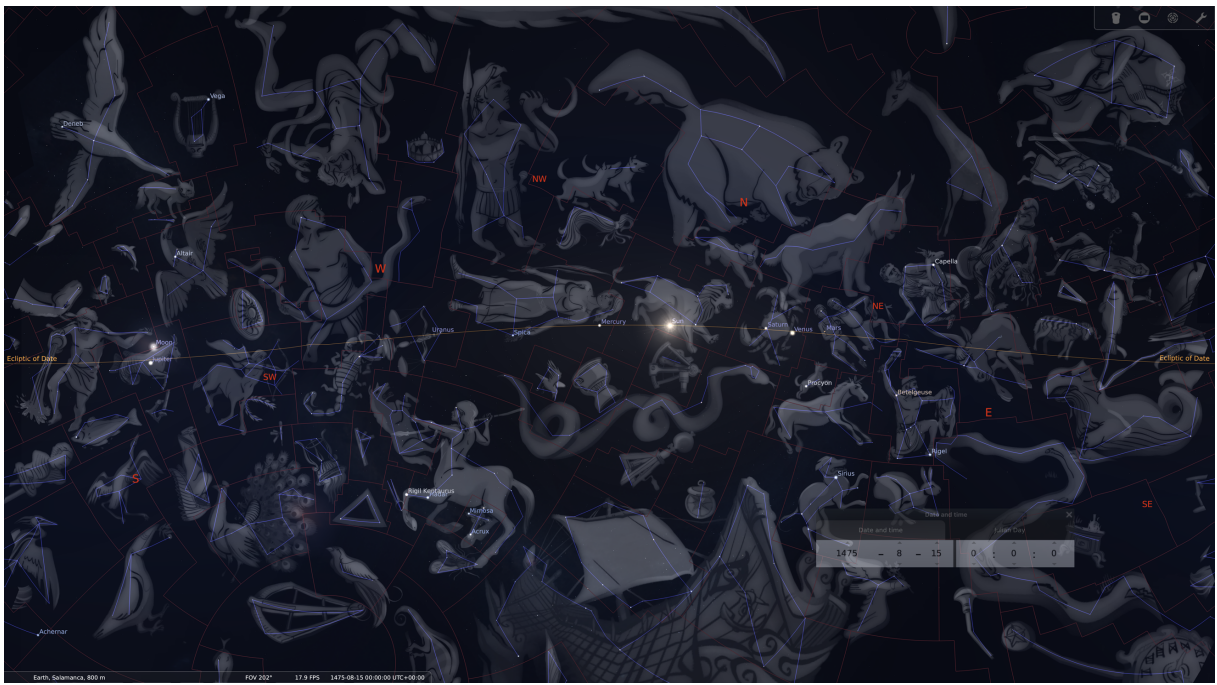

*Fig. 6. Celestial sphere plotted with Stellarium v0.21.3 in Mercator projection at 0:0:0 UTC, 15 August 1475.*

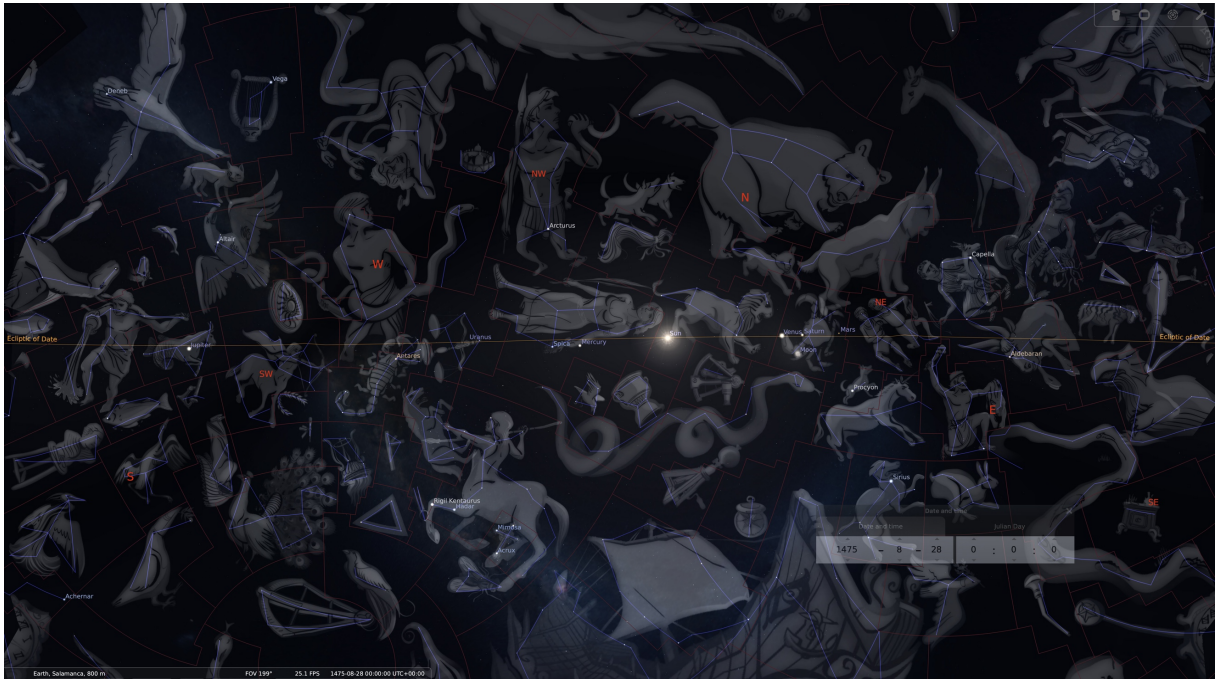

*Fig. 7. Celestial sphere plotted with Stellarium v0.21.3 in Mercator projection at 0:0:0 UTC, 28 August 1475.*

The next researcher to analyse the subject was Professor Esteban Lorente in 1993. In his article, [2, p. 315] he writes [our translation from Spanish]:

> *"There is a margin of a few days, between 15 and 28 August 1475, but at this time we have not found any event worthy of commemoration. We can be sure that this is the only possible year between 1450 and 1530; we have looked at the possibility of other important dates for the University, such as 1218, 1243, 1254 and 1422, but none of them can be represented here; the years 1467 and 1526 are somewhat close."*

We have seen above that the first statement is correct. We have also checked with Stellarium v0.21.3 that there are indeed no possible cases in the interval and the dates indicated as impossible. Doing the calculations for 1467 we have seen that it is possible from 23 to 26 August. However, as Fig. 8 shows, there are no feasible dates in 1526 because when Mercury enters Virgo on August 25, Mars is in Leo and remains in that constellation for the rest of the month.

The most recent work is that of Professor Hernández Pérez [4], published in 2021. In it she analyses the period 1460-1493 and tells us [4, p. 1443] that [our translations from Spanish]:

> *"... the only possible dates obtained are those between 20 and 24 August 1467 or between 13 and 24 August 1475."*

The year 1467 has been dealt with above when analysing the article by Professor Esteban Lorente. We simply recall here that the correct possible dates for that year are 23 to 26 August.

As for the year 1475, which we have considered when reviewing the work of Professor Noehles-Doerk, we reiterate that the feasible period is between 15 and 28 August. In fact, the 13th of that month is invalidated because, as shown in Fig. 9, the Moon was in Sagittarius.

On the other hand, Professor Hernández Pérez tells us [4, p. 1442] that:

> *"it is proposed as an initial hypothesis that the date chosen to determine the disposition of the stars was related to some singular event of an astronomical nature so that the pictorial* *representation of the celestial vault that was to cover the library would highlight the erudition of the team of astronomer professors at the University of Salamanca."*

*Fig. 8. Celestial sphere plotted with Stellarium v0.21.3 in Mercator projection at 0:0:0 UTC, 25 August 1526.*

*Fig. 9. Celestial sphere plotted with Stellarium v0.21.3 in Mercator projection at 0:0:0 UTC, 13 August 1475.*

She then states [4, 1444] that:

> *"The study has focused on the identification of conjunctions of two, three or more planets in the periods indicated. Thus, from 20 to 24 August 1467, Venus is positioned in Cancer, Mars in Pisces, Jupiter in Gemini and Saturn in Aries and therefore, there is no conjunction between any of these four planets. The situation is different between August 13 and 24, 1475 as Venus, Mars and Saturn are in the sign of Cancer and their positions are closer and equidistant between August 13 and 16. Conjunctions of two planets occur more frequently than those of three planets, and a conjunction of three planets in a constellation such as Cancer, which, as we shall see, was located in the central section of the vault and at the beginning of which the summer solstice, the day with the most hours of sunshine, occurs very rarely. All the attributes that an astronomer could consider relevant for the choice of a date that deserved to be perpetuated visually converged in this period."*

The following points should be made. By definition, a conjunction between two celestial bodies occurs when either their right ascensions or their ecliptic longitudes are equal. Even ignoring the incorrect dates just discussed and taking the correct ones seen above, in none of the periods specified for 1467 and 1475 does a triple conjunction occur between the planets indicated. But even if we admit the author's definition [4, p. 1443]:

> *"Two or more planets are in conjunction when they occupy very close positions in the night sky visible from the Earth."*

the situation for the day 15 August 1475 is not the one she has described above. One can see in Fig. 10, that Venus and Saturn are in Cancer, while Mars is in Gemini. According to the reasoning that the author has followed to rule out conjunctions in the year 1467, we must conclude that on the night we are analysing there is no conjunction either.

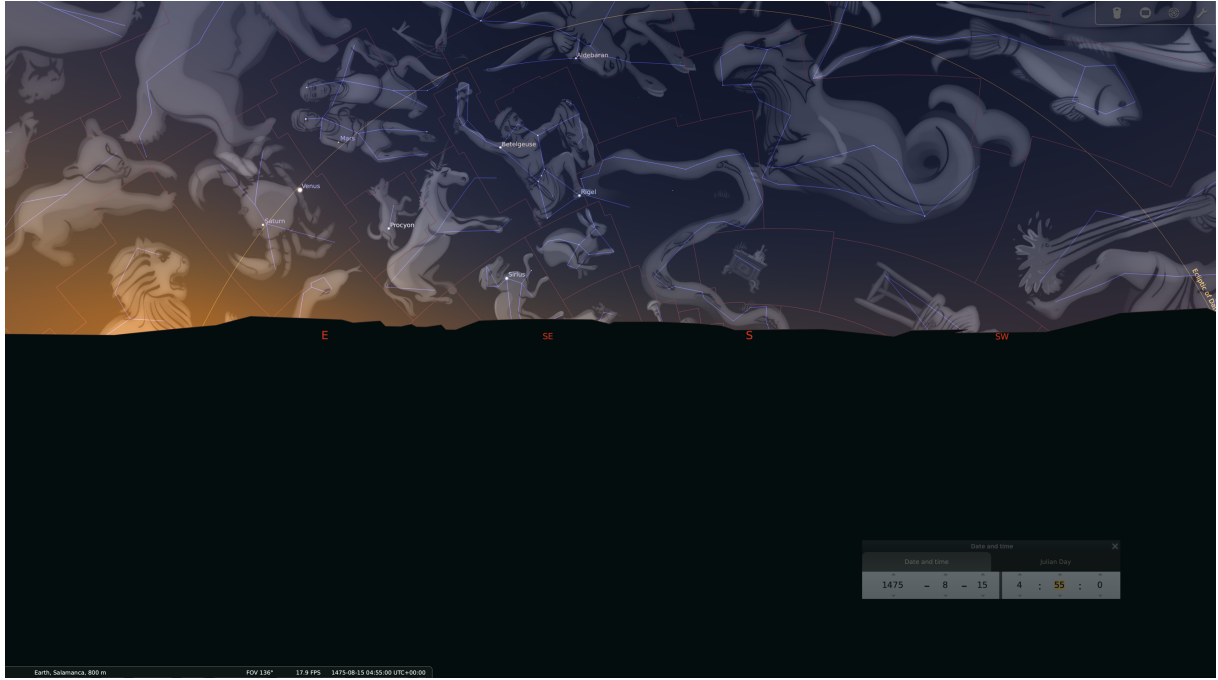

*Fig. 10. Positions of Saturn, Venus and Mars obtained with Stellarium v0.21.3 at 4:55:00 UTC, 15 August 1475.*

On the contrary, as shown in Fig. 11, on August 22, 1475, there is a conjunction, and a very close one (the angular separation is 41 minutes of arc), between Venus and Saturn in Cancer, with Mars still in Gemini.

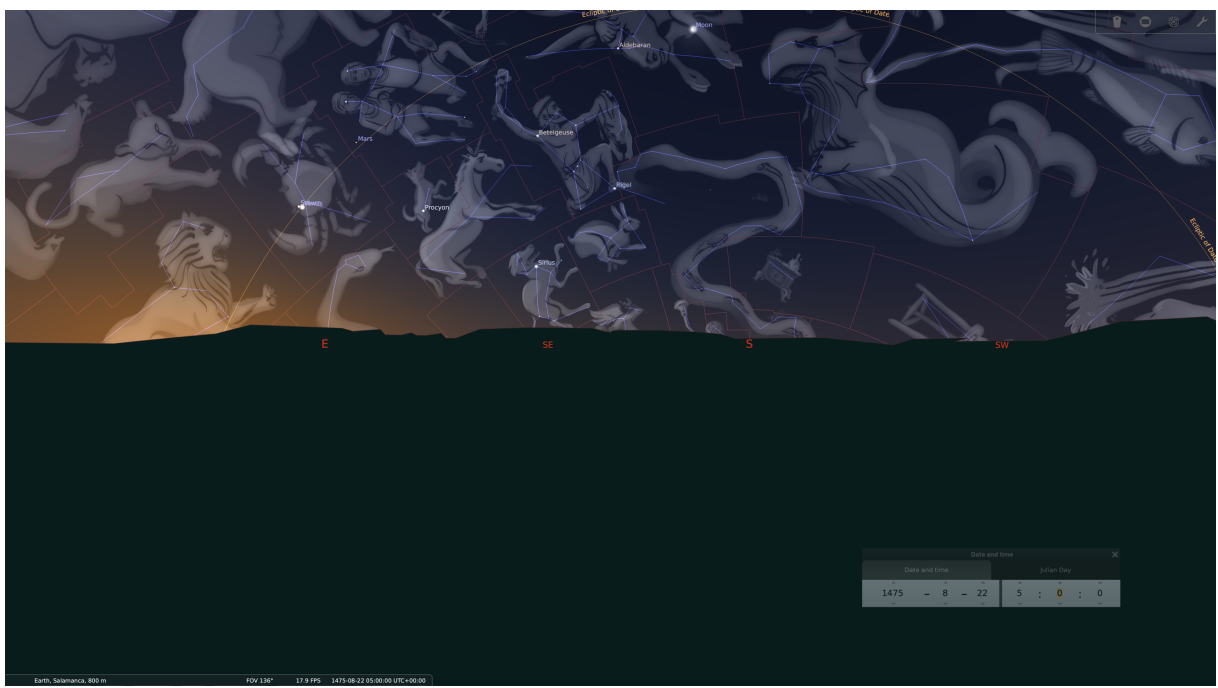

*Fig. 11. Positions of Saturn, Venus and Mars obtained with Stellarium v0.21.3 at 5:00:00 UTC, 22 August 1475.*

Finally, if we use the author's reasoning to identify a triple conjunction in Cancer, it would occur on 28 August 1475. Indeed, in Fig. 12, we see that on that day Venus, Saturn and Mars are in Cancer and their separation at that time is as symmetrical as possible since Mars entered that constellation on 25 August.

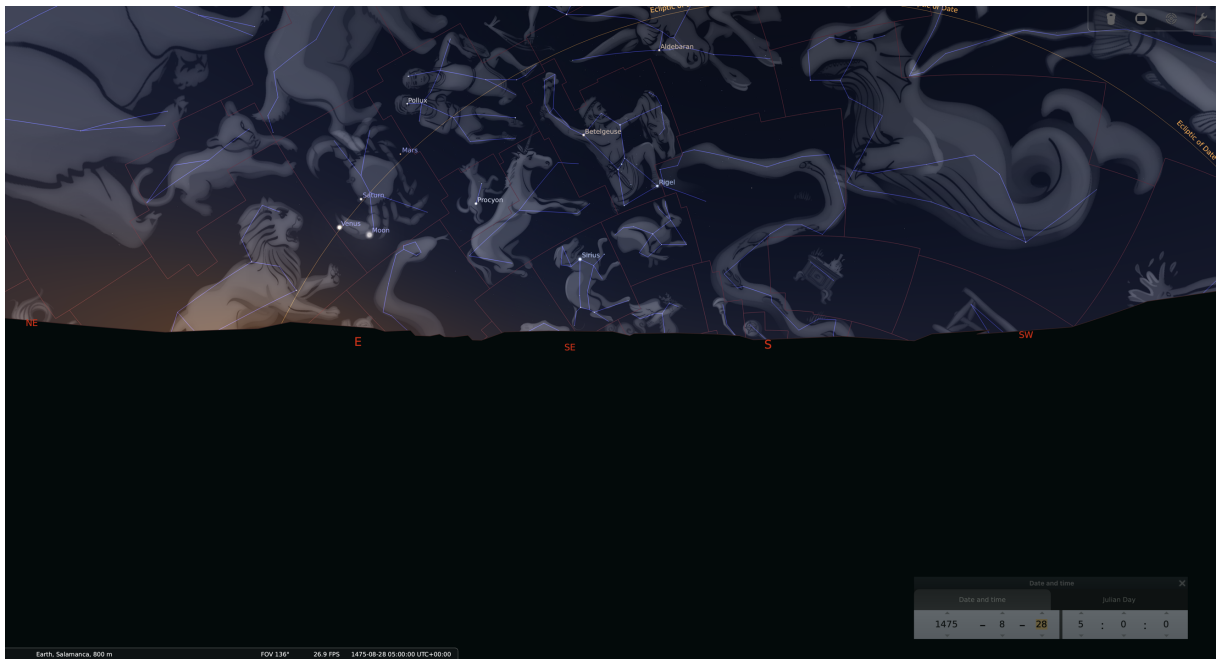

*Fig. 12. Positions of Saturn, Venus and Mars obtained with Stellarium v0.21.3 at 5:00:00 UTC, 28 August 1475.*

In view of the above analysis, it makes sense to say that a day is a "Cielo de Salamanca" day if at 0 hours UTC on that day the Sun is in Leo, Mercury is in Virgo and Venus, Mars, Jupiter, Saturn and the Moon are not in Leo, Virgo, Libra, Scorpio or Sagittarius. This arrangement of planets, Moon and Sun is called a "Cielo de Salamanca" (CdS) configuration. Note that this definition does not depend on the geographical location of the observer. Similarly, a year is said to be a "Cielo de Salamanca" year if it has "Cielo de Salamanca" days.

As a complement to the literature review, we have made a complete analysis using Stellarium v0.21.3 of the period between 1200 and 2300 to identify other years with "Cielo de Salamanca" days. We have chosen the beginning of the period before the University of Salamanca was founded in 1218. We show below in a table the results we have obtained.

| "Cielo de Salamanca" years and days | | | |
|---|---|---|---|
| Year | Interval | Days | Maximum angular separation Mercury-Sun during the interval |
| 1227 | 4 - 9 August | 6 | 26° 56' 37'' |
| 1240 | 3 - 14 August | 12 | 27° 21' 47'' |
| 1264 | 15 - 21 August | 7 | 18° 54' 05'' |
| 1443 | 8 - 23 August | 16 | 27° 07' 33'' |
| 1467 | 23 - 26 August | 4 | 11° 31' 23'' |
| 1475 | 15 - 28 August | 14 | 25° 55' 57'' |
| 1561 | 22 - 31 August | 10 | 26° 54' 57'' |
| 1582 | 6 - 16 August | 11 | 27° 07' 44'' |
| 1646 | 24 August - 7 September | 15 | 25° 51' 27'' |
| 1667 | 17 - 18 August | 2 | 27° 06' 38'' |
| 1678 | 30 August - 11 September | 13 | 20° 45' 18'' |
| 1785 | 19 August - 1 September | 14 | 27° 11' 24'' |
| 1798 | 24 August - 5 September | 13 | 27° 06' 48'' |
| 1857 | 1 - 5 September | 5 | 27° 00' 52'' |
| 1881 | 8 - 14 September | 7 | 11° 59' 54'' |
| 2022 | 22 - 25 August | 4 | 27° 15' 15'' |
| 2060 | 8 - 16 September | 9 | 26° 21' 01'' |
| 2092 | 13 - 17 September | 5 | 20° 22' 04'' |
| 2180 | 2 - 16 September | 15 | 25° 58' 36'' |
| 2199 | 3 - 17 September | 15 | 27° 05' 26'' |
| 2212 | 9 - 15 September | 7 | 27° 00' 00'' |
| 2263 | 16 - 20 September | 5 | 19° 02' 27'' |
| 2295 | 20 - 21 September | 2 | 09° 01' 48'' |

It is evident from this table that "Cielo de Salamanca" years are extremely infrequent. In the 1100-year period analysed there are only 23. We can see in Fig. 13 that there are extremely long intervals without possible configurations, with 179, 86, 64, 107, 59, 141 and 88 years. This is caused by the commensurability ratios between the orbital periods of the planets.

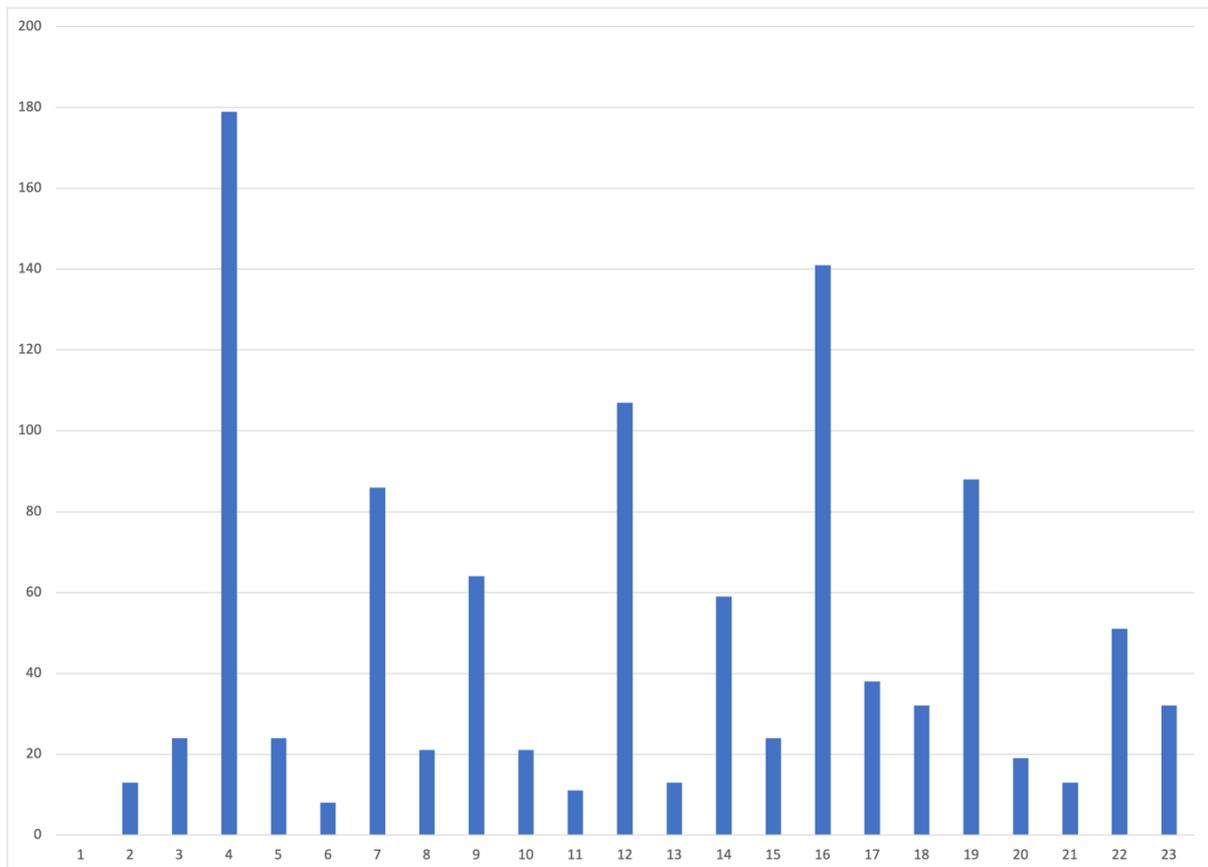

*Fig. 13. Elapsed intervals between years with "Cielo de Salamanca" days from 1200 to 2300.*

It is remarkable that 2022 is a "Cielo de Salamanca" year, that 141 years have passed since the previous one in 1881 and that for the next occasion we will have to wait until 2060. We propose to commemorate these events by celebrating the "Cielo de Salamanca" marathon, by ananlogy with the well-known Messier marathon. This new event will be held on those days when a CdS configuration is maintained from sunset to sunrise the following day. The marathon consists of observing Mercury after sunset, Mars, Jupiter and Saturn during the night, concluding with the observation of Venus just before dawn. Figs. 14, 15 and 16 show the three highlights of the "Cielo de Salamanca" marathon" to be held on 21 August 2022.

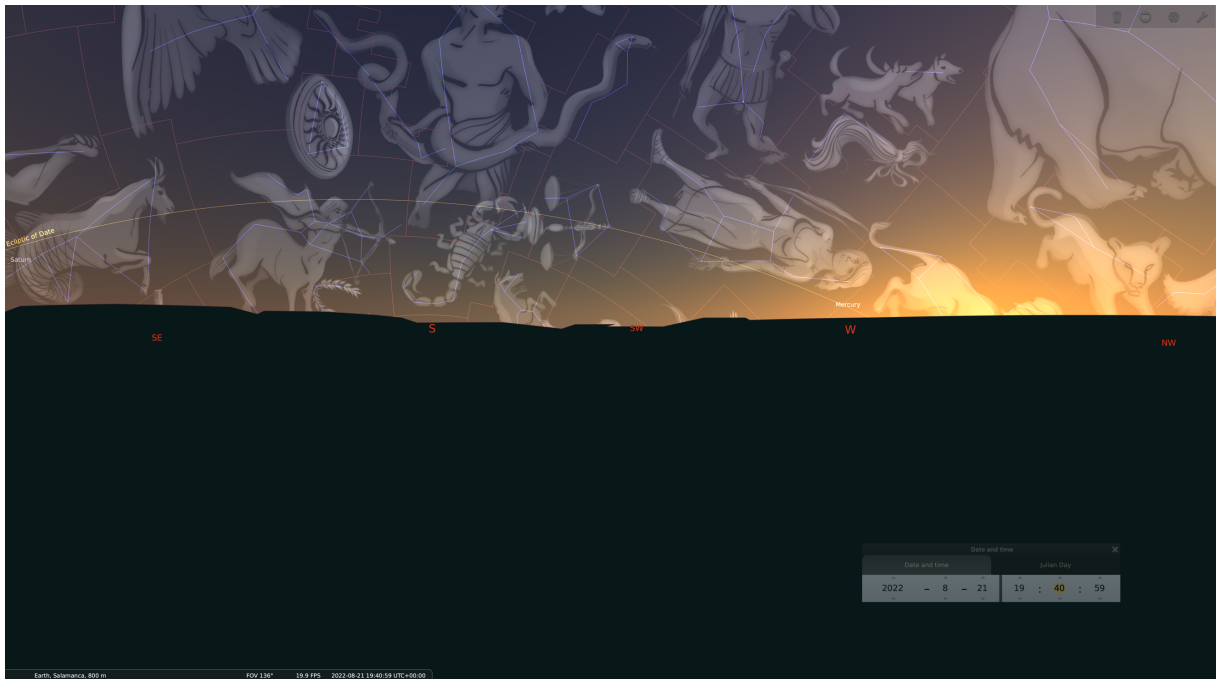

*Fig. 14. Results obtained with Stellarium v0.21.3 for sunset on 21 August 2022.*

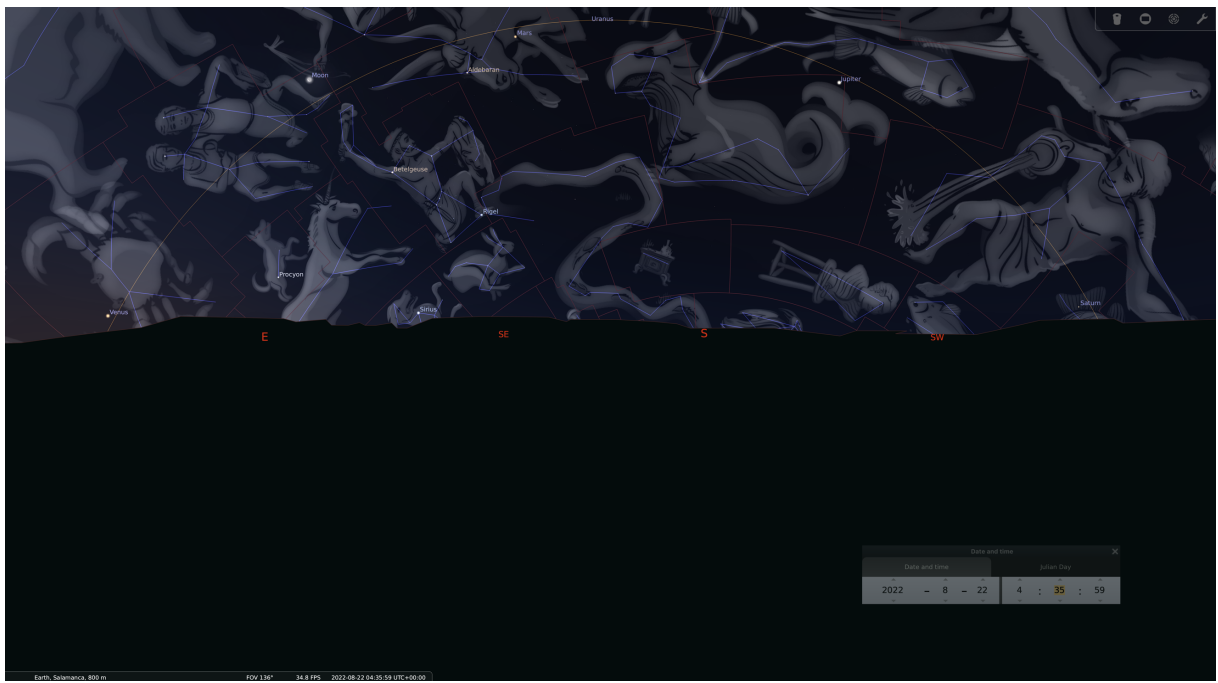

*Fig. 15. Results obtained with Stellarium v0.21.3 for the night of 22 August 2022.*

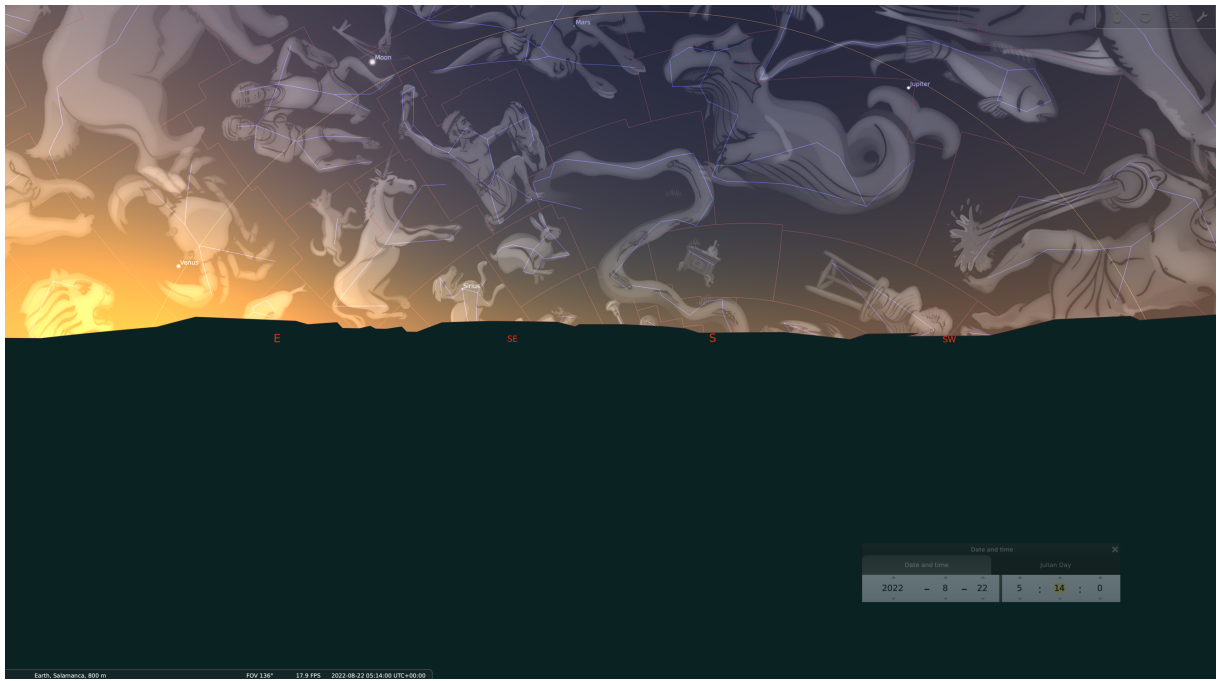

*Fig. 16. Results obtained with Stellarium v0.21.3 for sunrise on 22 August 2022.*

Several proposals for a complete reconstruction of the original vault of the first library of the University of Salamanca have appeared in the literature.

To begin with, we may consider in first place, the work of Recio Sánchez [9] of 2019 in which a very interesting proposal is made, as it does not depend on the assignment of a specific date to *El Cielo de Salamanca*. The author considers the astrological theory of planetary regencies over the zodiacal constellations. According to this, the Sun ruled over Leo, the Moon over Cancer, Venus over Taurus and Libra, Mars over Aries and Scorpio, Jupiter over Pisces and Sagittarius, and finally Saturn over Aquarius and Capricorn. He reasons that in the preserved part we see precisely the Sun next to Leo, which it rules, and Mercury is adjacent to Virgo, which is one of its rulerships. Then he hypothesises that the zodiacal constellations would be distributed in an arc over the vault in order to differentiate the southern hemisphere from the boreal hemisphere. The iconographic approach for the rest of the vault would be the same as in the preserved part. Therefore, the zodiacal constellations would be accompanied in parallel by their planetary regencies, which would occupy the boreal hemisphere because it would have a larger display area. This leads to a univocal spatial distribution that is coherent with the "Sky of Salamanca". Furthermore, this proposal leads to a homogeneous spatial distribution of the planets and constellations. This is in accordance with its aim of serving as a planisphere for teaching Astronomy. In any case, it remains to be explained why Mercury is precisely in Virgo and not in Gemini, and why there are no other planets in the preserved part of the vault.

Professor Hernández Pérez has outlined in her recent work [4] another proposal for a reconstruction based on the date she had assigned to the "Sky of Salamanca". However, as we have seen above, it is not possible to make a single dating, so that we should properly speak of a multiplicity of possible reconstructions. However, a circumstantial argument could be made for a reconstruction based on a specific lapse of time related to the period of construction of the ancient library on the grounds that, of the very rare years with possible configurations, one and only one of them, 1475, falls within that interval

These two cases and their associated difficulties serve to highlight that there could be other possibilities for the original configuration of the vault, which could be based on other considerations that we do not know of at this stage.

**The mystery of *El Cielo de Salamanca***

In view of the analysis of the results present in the literature and those obtained by us, our opinion is that, unless new evidence appears, the information contained in *El Cielo de Salamanca* is not sufficient to determine a single night. After all, it is a marvellous work of art that we can enjoy just by looking at it. The fact that there is a certain mystery surrounding it makes it, if possible, more interesting to the spectator's eyes. And what better complement than the unique opportunity we will have this year to observe live the magnificent spectacle of the "Sky of Salamanca" over several nights in August! We are sure that after contemplating it, we will continue to reflect on what the professors of Astrology at the University of Salamanca really wanted to tell us with the planetarium they designed at the end of the 15th century. A time of splendour of their science that was opening to the call of the Renaissance.